\documentclass[onecolumn,preprint,preprintnumbers,amsmath,amssymb,apsrev4-1,prb]{revtex4-1}
\usepackage{graphicx}
\usepackage{dcolumn}
\usepackage{bm}
\usepackage{SIunits}
\usepackage{verbatim}
\usepackage{placeins}
\usepackage{hyperref}

\begin{document}
\title{Topological magnons in one-dimensional ferromagnetic Su-Schrieffer-Heeger model with anisotropic interaction}
\author{Peng-Tao Wei$^{1,2}$}
\author{Jin-Yu Ni$^{1,2}$}
\author{Xia-Ming Zheng$^{3,1}$}
\author{Da-Yong Liu$^{3,1}$}
\email[]{dyliu@ntu.edu.cn}
\author{Liang-Jian Zou$^{1}$}
\email[]{zou@theory.issp.ac.cn}
\affiliation{1 Key Laboratory of Materials Physics, Institute of Solid State Physics, HFIPS, Chinese Academy of Sciences, Hefei 230031, China}
\affiliation{2 Science Island Branch of Graduate School, University of Science and Technology of China, Hefei 230026, China}
\affiliation{3 Department of Physics, School of Sciences, Nantong University, Nantong 226019, China}
\href{http://orcid.org/0000-0003-4370-473X}{ORCiD: 0000-0003-4370-473X}
\date{\today}
\begin{abstract}
Topological magnons in a one-dimensional (1D) ferromagnetic (FM) Su-Schrieffer-Heeger (SSH) model with anisotropic exchange interactions are investigated. Apart from the intercellular isotropic Heisenberg interaction, the intercellular anisotropic exchange interactions, {\it i.e.} Dzyaloshinskii-Moriya interaction (DMI) and pseudo-dipolar interaction (PDI), also can induce the emergence of the non-trivial phase with two degenerate in-gap edge states separately localized at the two ends of the 1D chain, while the intracellular interactions instead unfavors the topological phase. The interplay among them has synergistic effects on the topological phase transition, very different from that in the two-dimensional (2D) ferromagnet. These results demonstrate that the 1D magnons possess rich topological phase diagrams distinctly different from the electronic version of the SSH model and even the 2D magnons. Due to the lower dimensional structural characteristics of this 1D topological magnonic system, the magnonic crystals can be constructed from bottom to top, which has important potential applications in the design of novel magnonic devices.
\end{abstract}

\vskip 300 pt

\maketitle

\section{Introduction}
Since the discovery of quantum Hall effect \cite{PRL45-494,PRL48-1559,PRL49-405,RMP58-519}, the topological phase of matter has attracted extensive attention. Although the topological phase of matter is first found in electronic systems, it is later extended to the field of quasi-particles, such as phonons \cite{PRL105-225901}, photons \cite{PRL100-013904,NM12-233} and even magnons \cite{Sci329-297}, {\it etc}. Katsura {\it et al.} \cite{PRL104-066403} theoretically predicted the thermal Hall effect of magnons. Subsequently, Onose {\it et al.} \cite{Sci329-297} experimentally confirms the existence of magnon Hall effect in Lu$_{2}$V$_{2}$O$_{7}$ with pyrochlore structure, which sets off an upsurge of studying the topological properties of magnons. So far, various non-trivial topological phases of magnons for different crystal structures and magnetic interactions had been presented. In two-dimensional (2D) system, it is found that the anisotropic magnetic interaction, such as the magnetic dipole-dipole interaction \cite{PRB87-174402,PRB90-104417,PRB98-224409}, Dzyaloshinskii-Moriya interaction (DMI) \cite{PRB97-174413,PRB99-214424,PRB99-174412,JPCM29-185801} and pseudo-dipolar interaction (PDI) \cite{PRB95-014435,PRApp9-024029} rather than the isotropic magnetic interaction (Heisenberg-type exchange interaction), usually produce the non-trivial topological phase of magnon in different types of lattices, {\it e.g.} honeycomb lattice, Kagome lattice and triangular lattice, {\it etc}. These anisotropic interactions often open an energy gap and lead to topological magnon insulator, and can be seen as effective spin-orbit couplings in electronic topological insulator. Due to the inherent topological protection, the topological magnons generally have a wider application prospect than the traditional ones.
Due to the abundance of magnetic materials and magnetic structures \cite{EPL103-47010}, as well as the manipulation of various magnetic interactions, magnons have become an important platform for topological physics and quantum technology \cite{PR915-1,ARCMP13-171}.
As a consequence, it has great potential application for designing novel quantum devices in magnonics \cite{JPD43-264001,NP11-453}.

As we know, there also are abundant topological phases in one-dimensional (1D) electronic systems, {\it e.g.} Su-Schrieffer-Heeger (SSH) model and Kitaev chain, {\it etc}. Indeed unlike electrons, the magnon topological phases are easier to be realized by regulating magnetic interactions. Although the 2D magnons are extensively studied in previous years and numerous topological magnon states ({\it e.g.} magnonic topological insulator, magnonic topological semimetal with Dirac or Weyl magnon, {\it etc}.) are discovered, the topological properties of magnons in 1D system remain unclear. We can expect more abundant topological phases in 1D magnon systems than in electronic systems due to the rich regulation of magnetic interactions. We notice that recently Pirmoradian {\it et al.} \cite{PRB98-224409} constructed a 1D magnetosphere chain \cite{PRL42-1698}, in which there is only dipole-dipole interaction rather than Heisenberg-type interaction between macrospins on the magnetospheres, and they discovered topological magnons by manipulating the external magnetic field. As a matter of fact, the magnetosphere chain they studied is a classical magnetic macrosystem. However, the mechanism to induce intrinsic topological magnons without external field is still lacking, especially the topological properties in the lattice model of the realistic magnetic materials is deserved to explore.

In this paper, here we investigate the possible magnonic topological phases and the related topological phase transitions in a 1D ferromagnetic (FM) SSH model by manipulating both the isotropic Heisenberg exchange interaction and the anisotropic exchange interactions, including DMI and PDI. To characterize the topological phase, we adopt the Zak phase for the infinite system and the real-space topological number for the finite system, respectively. We first investigate the pure Heisenberg-type interaction, and find that the topological phase can be realized by adjusting the ratio of intracellular interaction $J_{1}$ and intercellular interaction $J_{2}$, {\it i.e.} $J_{2}$$/$$J_{1}$. This result is very different from that in the 2D honeycomb ferromagnet, in which there is no topological phase transition for only isotropic Heisenberg-type interaction. Moreover, the introduction of anisotropic interactions, {\it {\it i.e.}} DMI (intracellular interaction $D_{1}$ and intercellular interaction $D_{2}$) and PDI (intracellular interaction $F_{1}$ and intercellular interaction $F_{2}$), also plays an important role in the realization of the topological phases. We demonstrate that the intercellular interactions ($J_{2}$, $D_{2}$, and $F_{2}$) favor topological non-trivial phase, while the intracellular interactions ($J_{1}$, $D_{1}$, and $F_{1}$) hinder it, in the 1D FM SSH model. This paper is organized as follows: the theoretical model and method are introduced in Sec. II; the numerical results and discussions are presented in Sec. III; and the last section is the remarks and conclusions.

\section{Model and Method}
\begin{figure}[htbp]
\hspace*{-2mm}
\centering
\includegraphics[trim = 0mm 0mm 0mm 0mm, clip=true, angle=0, width=1.0 \columnwidth]{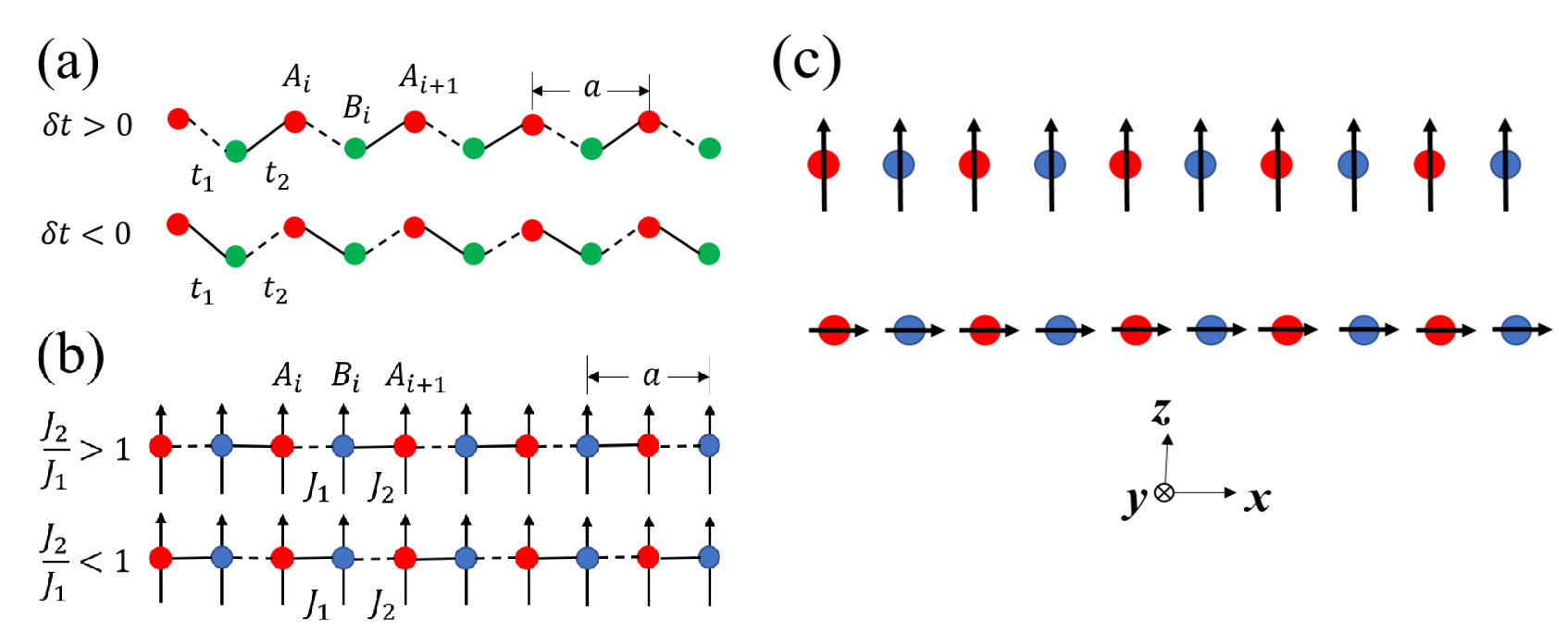}
\caption{(Color online) (a) 1D electronic SSH model, (b) 1D FM SSH model. Each unit cell contains two lattice sites A and B. The solid and dotted lines indicate the strong and weak exchange interactions, respectively. $J_{1}$ and $J_{2}$ denote the intracellular and intercellular interactions, respectively. (c) Two types of FM ground states: the spin quantization axes are $z$-axis (top panel) and $x$-axis (bottom panel), respectively.}
\label{Fig1}
\end{figure}

Analogous to the electronic SSH model \cite{PRL42-1698}, as shown in Fig.~\ref{Fig1}(a), we construct a 1D FM SSH model including the intracellular and intercellular interactions, {\it i.e.} $J_{1}$ and $J_{2}$, which presents a pure Heisenberg-type model. In general, there are two types of FM ground states, that is, the spin orientation is perpendicular to or along the direction of the 1D chain, as shown in Fig. ~\ref{Fig1}(c), and thus these two quantization axes are considered, respectively.

The Hamiltonian of the 1D spin chain with anisotropic exchange interactions can be written as
\begin{eqnarray}
  H&=&H_{0}+H_{AE}+H_{DM}+H_{PD}.
  \label{Eq1}
\end{eqnarray}
$H_{0}$ is the Heisenberg exchange interaction, reading
\begin{eqnarray}
  H_{0}&=&-J_{1}\sum_{i=1}^{N}{\mathbf{S}_{i,A}\cdot \mathbf{S}_{i,B}}-J_{2}\sum_{i=1}^{N-1}{\mathbf{S}_{i,B}\cdot \mathbf{S}_{i+1,A}},
  \label{Eq2}
\end{eqnarray}
where $J_{1}$ and $J_{2}$ represent the exchange interaction in the unit cell (intracellular) and between the two unit cells (intercellular), respectively. And $H_{AE}=-\frac{1}{2}\sum_{i}{K_{i}{{(S}_{i}^{z})}^{2}}$
is the easy-axial anisotropy term for the quantization $z$-axis, where $K_{i}$ is the axial anisotropy energy ($K_{i}$=10$J_{1}$ is used for all the calculations). $H_{DM}$ denotes the DMI term \cite{JPCS4-241,PR120-91}, originating from the spin-orbit coupling, while $H_{PD}$ represents the PDI \cite{PR52-1178,PRL102-017205}, mainly contributed from both the strong spin-orbit coupling and the orbital degree of freedom.

The DMI and PDI are covered by the following forms:
\begin{eqnarray}
H_{DM}&=&\mathbf{D}_{1}\cdot\sum_{i}\left(\mathbf{S}_{i,A}\times \mathbf{S}_{i,B}\right)\\ \nonumber
&&+\mathbf{D}_{2}\cdot\sum_{i,i+1}\left( \mathbf{S}_{i,B}\times\mathbf{S}_{i+1,A}\right),
\label{Eq3}
\end{eqnarray}
\begin{eqnarray}
H_{PD}&=&-F_{1}\sum_{i=1}^{N}{\left(\mathbf{S}_{i,A}\cdot \mathbf{e}_{i,AB}\right)\left(\mathbf{S}_{i,B}\cdot\mathbf{e}_{i,AB} \right)}\\ \nonumber
&&-F_{2}\sum_{i=1}^{N-1}{\left(\mathbf{S}_{i,B}\cdot \mathbf{e}_{i,i+1,BA}\right)\left(\mathbf{S}_{i+1,A}\cdot \mathbf{e}_{i,i+1,BA}\right)},
  \label{Eq4}
\end{eqnarray}
where $\mathbf{D}_{1}$ ($F_{1}$) and $\mathbf{D}_{2}$ ($F_{2}$) represent intracellular and intercellular nearest neighbor DMIs (PDIs), respectively. $\mathbf{e}_{i,AB}$ is the unit vector between lattice sites A and B in the $i$-th unit cell, and $\mathbf{e}_{i,i+1,BA}$ is the unit vector between lattice site B of the $i$-th unit cell and lattice site A of the $(i+1)$-th unit cell. $\mathbf{e}_{i,AB}$ and $\mathbf{e}_{i,i+1,BA}$ point in the direction of the chain.

We firstly analyze the magnetic ground state before calculating the magnon spectrum. The Heisenberg term with $-J<0$ favors the FM state, and the easy-axial anisotropy term with $K_{z}$ tends to align the spins parallel along the $z$ axis. However, in the presence of anisotropic magnetic exchange interactions, the nearest-neighboring DMI and PDI, the magnetic ground state may change. In the classical spin limit, the derivation process with respect to the spin canting angles \cite{PRB97-094412} is performed to obtain the minimal energy of the total Hamiltonian $H$ (Eq.(\ref{Eq1})). Two possible ground states are found to be the FM states along $z$-axis and $x$-axis with the total energies per unit cell $-(J_{1}+J_{2})-K$ and $-(J_{1}+J_{2})-(F_{1}+F_{2})$, respectively. Hence the phase boundary is $K=F_{1}+F_{2}$, that is, when $K>F_{1}+F_{2}$, it is FM state with spins oriented along the $z$ axis ($z$-FM phase), otherwise, it is FM state with the spins oriented along the $x$ axis ($x$-FM phase). Through the analysis of the total energy, it is found that the component of the DMI $D_{z}$, will not affect the $z$-FM phase because it does not contribute to the total energy. While the PDI $F>0$ prefers each pair of spins to be parallel to the bond between them ($x$ axis), so it can change the magnetic ground state. Note that all the parameter values used ($K=10J_{1}$, $F=J_{1}$, $0<F_{2}/F_{1}\leq 5$, for the convenience of theoretical analysis) fall into the $z$-FM phase, even for the large values of $D_{z}$ or $F$.

By the linear Holstein-Primakoff (HP) transformation \cite{PR58-1098} with bosonic generation (annihilation) operator (for A sublattice, $a^{\dag}$ ($a$); for B sublattice, $b^{\dag}$ ($b$)), the Hamiltonian of magnon in momentum $k$-space can be expressed as $H=\frac{1}{2}\sum_{k}{\Psi_{k}^{\dag}H_{k}\Psi_{k}}$, where $\Psi_{k}=\left(a_{k},a_{-k}^{\dag},b_{k},b_{-k}^{\dag}\right)^{T}$. Using the linear HP transformation, the spin operators for the A (or B) sublattice are given by
\begin{equation}
\begin{cases}
S_{A}^{\dag}=\sqrt{2S}a \\ \nonumber
S_{A}^{-}=\sqrt{2S}a^{\dag} \\ \nonumber
S_{A}^{z}=S-a^{\dag}a. \\
\end{cases}
\end{equation}

By the Bogoliubov transformation \cite{INC7-794}, we can diagonalize the magnon Hamiltonian and obtain the energy spectrum of the magnon. The magnon Hamiltonian $H_{k}$ satisfies the generalized eigenvalue problem \cite{PR139-A450},
\begin{eqnarray}
\eta H_{k}\Gamma_{k}&=&\Gamma_{k}\eta E_{k},
  \label{Eq5}
\end{eqnarray}
where $\Gamma_{k}$ is the transformation matrix that diagonalizes the
magnon Hamiltonian, $\eta$ is a metric matrix,
$\eta=I_{2 \times 2} \otimes \sigma_{z}$, $I$ is the identity
matrix, and $\sigma_{z}=\begin{pmatrix}
1 & 0 \\
0 & -1
\end{pmatrix}$ is the Pauli matrix. In the generalized eigenvalue
problem of boson, $E(k)=-E(-k)$ is artificially introduced. As a
result, we only need to consider the non$-$negative eigenvalues of the
energy spectrum. On the other hand, for a finite 1D system with $N$
unit cells, the generalized eigenvalue problem in real space is
$\eta H\Gamma=\Gamma\eta E$, where $H$ is the $2N \times 2N$
real space matrix, and $\eta=I_{N \times N} \otimes \sigma_{z}$.

As is known that, the electronic 1D SSH model possesses the chiral ($\Gamma$), spatial ($P$) and temporal ($T$) inversion symmetries \cite{PE119-113973} defined as $\Gamma H(k)\Gamma^{-1}=-H(k)$, $PH(k)P^{-1}=H(-k)$, $TH(k)T^{-1}=H(-k)$, respectively, in which
$\Gamma=I_{2 \times 2} \otimes \sigma_{z}$, $P=I_{2 \times 2} \otimes \sigma_{x}$, $T=I_{2 \times 2} \otimes \begin{pmatrix}
0 & i \\
i & 0
\end{pmatrix}$. The chiral symmetry requires on-site energy is 0, while
the spatial and temporal inversion symmetry only require the identical
on-site energies \cite{PRL127-147401}. In consequence, for the 1D FM SSH model with pure Heisenberg-type interaction, the chiral symmetry is broken, while the spatial and temporal inversion symmetries are preserved. When the anisotropic DMI and PDI exist in the system, the spatial inversion symmetry is broken while only the temporal inversion symmetry is preserved. Consequentially, the 1D FM SSH model of magnons has no direct correspondence to the electronic systems because of the different symmetries; thus, the topological magnons of the magnetic SSH model are not superficial extensions of the electronic SSH system. The most important difference is that the topological magnons involving anisotropic interaction features do not have direct correspondence to any electronic systems, though the isotropic Heisenberg term ($J_{1}$ and $J_{2}$) corresponds to the electronic interactions ($t_{1}$ and $t_{2}$). In addition, it should be pointed out that the change of quantization axis does not affect the symmetry of the system.

For a 1D infinite system, Zak phase \cite{PRL62-2747} is often used to describe its topological properties within the momentum $k$-space,
\begin{eqnarray}
\varphi_{Zak}&=&-i\int_{BZ}^{\ }{\Psi_{k}^{\dag}\frac{d}{dk}\Psi_{k}{dk}}.
  \label{Eq6}
\end{eqnarray}
When $\varphi_{Zak}$ is not 0, the topological phase is non-trivial,
and the system has topological edge state; otherwise, the topological
number is 0, and the topological phase is trivial, which means no
topological edge state in the system.

For a 1D finite system, we extend the real-space topological number of
the electronic (fermionic) version \cite{AP356-383} to the magnonic (bosonic) case. The
topological number of real space $N_{R}$ is described as
\begin{eqnarray}
N_{R}&=&\frac{1}{2}Sig((X+iH)\eta),
  \label{Eq7}
\end{eqnarray}
where $X$ is a matrix composed of rescaled coordinates consistent with
the size of $H$. For the $i$-th unit cell, its reduced coordinates are
$X_{i,j}=\frac{i\delta_{ij}}{N}$. $Sig(Q)$ is equal to the number
of positive eigenvalues of $Q=(X+iH)\eta$.

\section{Results and Discussions}
\subsection{Heisenberg-type interaction}
We first investigate the 1D FM SSH model with only pure Heisenberg-type isotropic exchange interaction. Through the linear HP transformation mentioned above, the Heisenberg interaction term $H_{0}$ can be expressed as
\begin{eqnarray}
H_{0}&=&\left(J_{1}+J_{2}+K \right)S\sum_{i} \left(a_{i}^{\dag}a_{i}+ b_{i}^{\dag}b_{i} \right)\\ \nonumber
&&-J_{1}S\sum_{i} \left(a_{i}b_{i}^{\dag}+ a_{i}^{\dag}b_{i} \right)
-J_{2}S\sum_{i} \left(b_{i}a_{i+1}^{\dag}+ b_{i}^{\dag}a_{i+1} \right).
  \label{Eq8}
\end{eqnarray}
The matrix of $H_{0}$ can be seen in the Appendix.

\begin{figure}[htbp]
\hspace*{-2mm}
\centering
\includegraphics[trim = 0mm 0mm 0mm 0mm, clip=true, angle=0, width=0.8 \columnwidth]{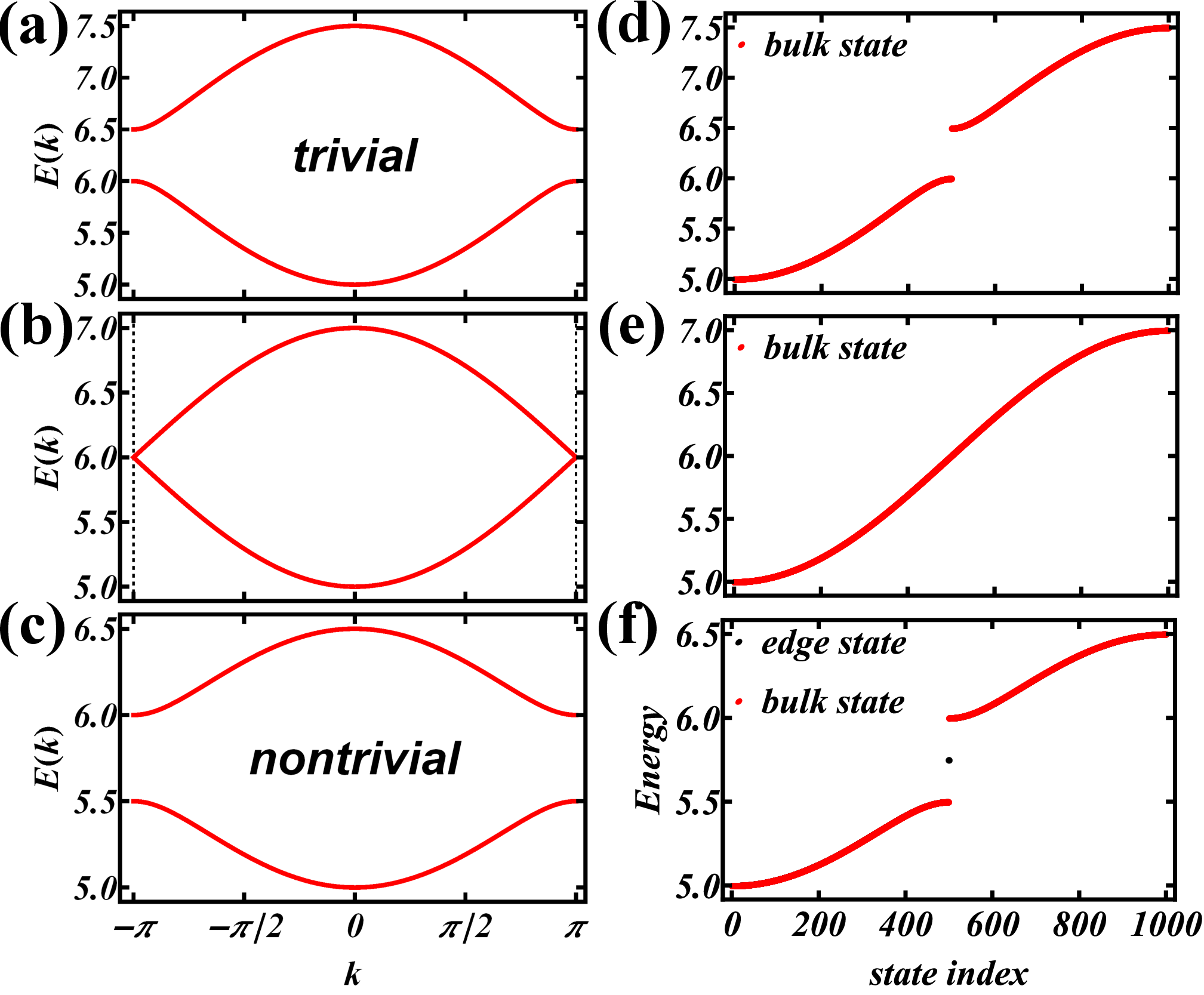}
\caption{(Color online) Band structures dependent on the ratio $J_{1}/J_{2}$ for 1D FM SSH model and the real-space energy spectrum of finite 1000 lattice sites of 1D chain, respectively: $J_{1}/J_{2}$=1.5 (a) and (d), 1.0 (b) and (e), 0.5 (c) and (f).}
\label{Fig2}
\end{figure}

The band structures dependent on the ratio $J_{1}/J_{2}$ are presented in Fig.~\ref{Fig2}(a)-(c).
It is obviously found that a topological phase transition occurs at a critical ratio $J_{1}/J_{2}$=1, accompanied by an energy gap close. This is very similar to the case of the electronic 1D SSH model \cite{TI2017}, that is the phase transition between topologically trivial and non-trivial phases can be realized by regulating the ratio of the intracellular and intercellular interactions, distinctly different from that in the 2D honeycomb ferromagnet in which the topological phase transition is mainly induced by the anisotropic interaction rather than the isotropic interaction \cite{JPCM29-185801}.
Notice that for the critical ratio $J_{1}/J_{2}$=1, the system exhibits a linear Dirac-like dispersion, but it is trivial owning to the band accidental degeneracy. Therefore, for $J_{1}/J_{2}$$\neq$1, although the energy band opens a gap, it is not necessarily topologically non-trivial.
It is demonstrated that the intercellular (intracellular) interaction $J_{2}$ ($J_{1}$) favors (unfavors) the topological non-trivial phase. Indeed, for $J_{1}<J_{2}$ (in comparison with $\delta t=t_{1}-t_{2}>0$ for the electronic SSH model), the magnon state is a 1D topological magnon insulator, as shown in Fig.~\ref{Fig2}(c). In addition, we note that the position of the gap close at $k=\pi$ or $-\pi$ is different from that of the electronic SSH model at $k=0$.
Owning to the finite energy gap of 1D topological magnon insulator, a topologically protected edge state can be expected according to the bulk-edge correspondence \cite{PRB95-035421}. In order to investigate the edge state of this topological magnon insulator, the energy spectrums of finite lattice sites in real space are also given, as displayed in Fig.~\ref{Fig2}(d)-(f). Accordingly, two degenerate edge states are found located within the band gap for the topological magnon insulating phase, which can be seen in Fig.~\ref{Fig2}(f).

\begin{figure}[htbp]
\hspace*{-2mm}
\centering
\includegraphics[trim = 0mm 0mm 0mm 0mm, clip=true, angle=0, width=0.8 \columnwidth]{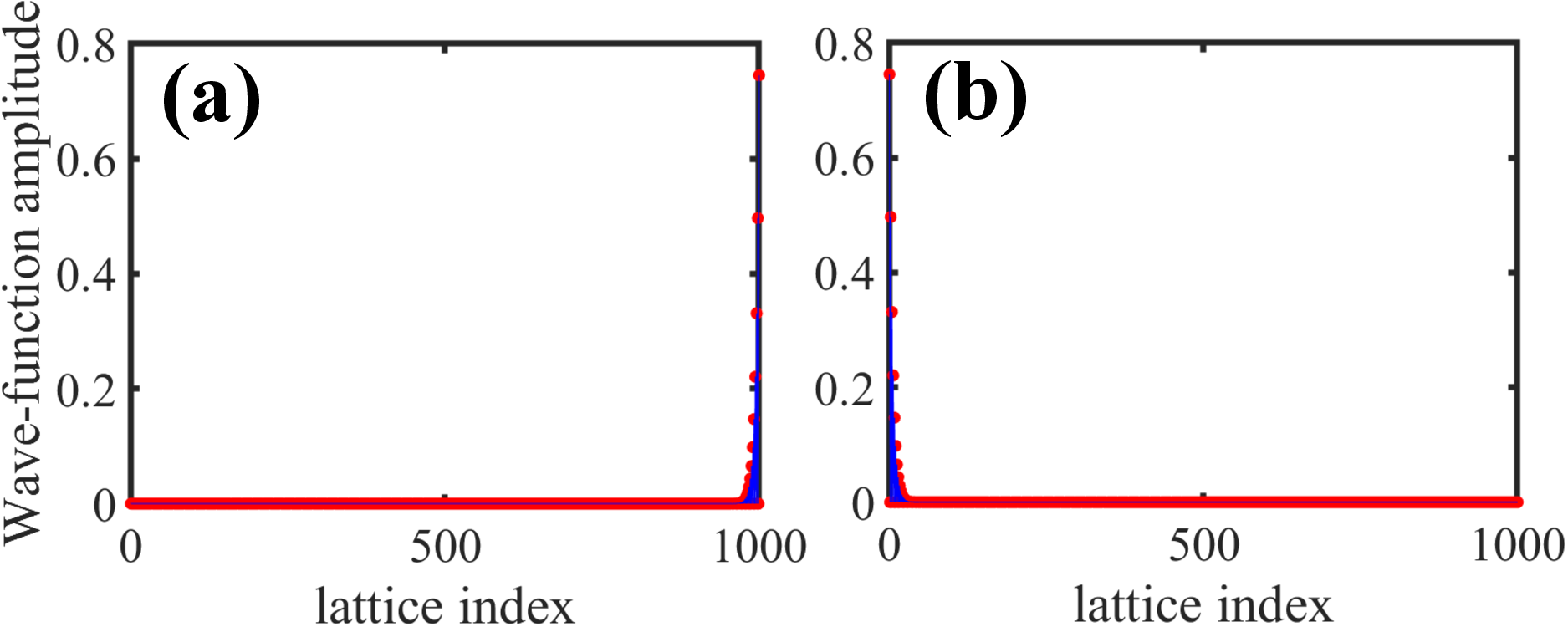}
\caption{(Color online) Wave-function amplitude dependent on the lattice index. The two degenerate edge states in Fig. 2(f) are separately localized at both ends of the 1D chain.}
\label{Fig3}
\end{figure}

Through the analysis of the real-space-resolved wave functions for the two degenerate edge states, it is found that the two edge states are separately localized at the two ends of the 1D chain, as seen in Fig.~\ref{Fig3}. In fact, this degenerate edge states can be simply lifted by introducing an inequivalence between A and B sites in the SSH model.
Considering the topological characteristic of this in-gap edge states, it can be regarded as zero-energy mode of magnon at finite frequency in 1D quantum magnets, which is very like the Majorana zero mode in 1D chain (nanowire) of electronic system \cite{Science336-1003}. It should be emphasized that the in-gap edge state as a novel type of bound state is distinctly different from the two-magnon bound state in 1D Heisenberg spin chain \cite{Nature502-76}.

\begin{figure}[htbp]
\hspace*{-2mm}
\centering
\includegraphics[trim = 0mm 0mm 0mm 0mm, clip=true, angle=0, width=0.6 \columnwidth]{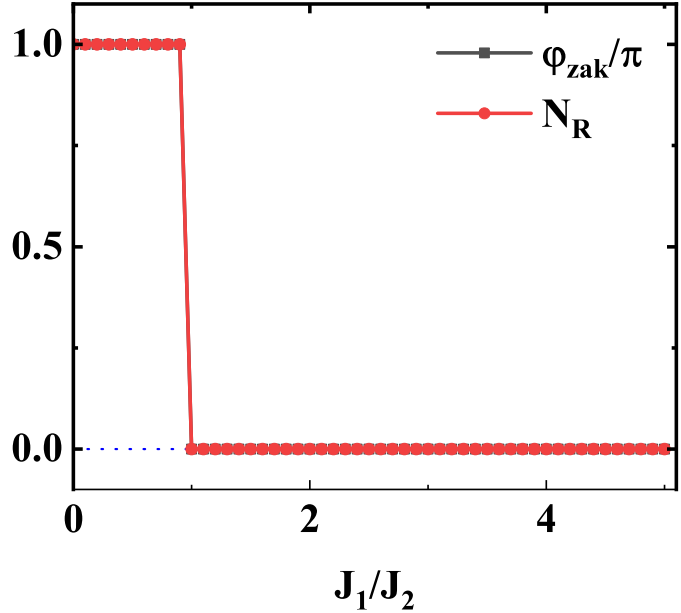}
\caption{(Color online) Momentum $k$-space Zak phase $\varphi_{Zak}$ and real-space topological number $N_{R}$ in the 1D FM SSH model with the Heisenberg-type interaction.}
\label{Fig4}
\end{figure}

The corresponding topological invariants in both the momentum $k$-space (Zak phase $\varphi_{Zak}$) and the real space ($N_{R}$) are shown in Fig.~\ref{Fig4}. It can be found that the topological number changes from 0 to 1 at the critical point $J_{1}$=$J_{2}$ after the energy gap is closed and reopened, indicating the phase transition process between the topological trivial and non-trivial phases. Notice that it can be seen that the topological number of $k$-space and real space are consistent with each other in describing the topological phase transition, implying our methods in calculating the topological invariants are effective.

For the quantization $x$-axis, as displayed in Fig.~\ref{Fig1}(c), the Heisenberg interaction term $H_{0}^{'}$ can be rewritten as
\begin{eqnarray}
H_{0}^{'}&=&\left(J_{1}+J_{2} \right)S\sum_{i} \left(a_{i}^{\dag}a_{i}+ b_{i}^{\dag}b_{i} \right)\\ \nonumber
&&-J_{1}S\sum_{i} \left(a_{i}^{\dag}b_{i}+h.c. \right)
-J_{2}S\sum_{i} \left(b_{i}^{\dag}a_{i+1}+h.c. \right)
  \label{Eq9}
\end{eqnarray}
Obviously, it is consistent with the form of the quantization $z$-axis which indicates that the change of the quantization axis does not affect our results for the pure Heisenberg-type interaction.

\subsection{In the presence of DMI}
Although the isotropic Heisenberg interaction can induce the topological phase in 1D spin chain, the anisotropic interactions play an essential role in topological phase transition in 2D lattices, such as honeycomb \cite{PRL117-227201,PRB95-014435,PRB97-174413,PRB99-214424} and Kagome \cite{PRB87-144101,PRB90-024412} lattices, {\it etc}. Therefore the influence of the anisotropic exchange interactions deserves investigation. Here we first focus on the DMI originating from the spin-orbit coupling. Using the linear HP transformation, the DMI term $H_{DM}$ (more details are shown in Appendix) can be written as
\begin{eqnarray}
H_{DM}&=&-iD_{1z}S\sum_{i}\left(a_{i}^{\dag}b_{i}-a_{i}b_{i}^{\dag} \right)\\ \nonumber
&&-iD_{2z}S\sum_{i,i+1}\left(b_{i}^{\dag}a_{i+1}-b_{i}a_{i+1}^{\dag} \right),
  \label{Eq10}
\end{eqnarray}
where $D_{1z}$ and $D_{2z}$ are the components of intracellular
$\mathbf{D}_{1}$ and intercellular $\mathbf{D}_{2}$ on the
quantization $z$$-$axis of spin, respectively. For simplicity, we fix
$D_{1z}=J_{2}=1$ for the $J_{1}/J_{2}$ case and $D_{1z}=J_{1}=1$ for the $J_{2}/J_{1}$ case in all the calculations. Note that due to the chiral characteristic of DMI, the values of both $D_{1z}$ and $D_{2z}$ can be positive or negative for 1D chain,
and all cases of four possible signs are considered. It is found that the change of sign does not change the topological phase diagram, but only the position of the closing gap in the energy spectrum during the phase transition. For $D_{2z}/D_{1z}>0$, the energy gap closes at $k=\pi/2$ (Note that for the pure DMI alone, the gap closes at $k=0$, not shown here, different from the case of the pure Heisenberg interaction.); otherwise, the position where energy gap closes is associated with $D_{2z}/D_{1z}$. Therefore, we only consider two cases with the relative sign of the $D_{2z}/D_{1z}$ being positive or negative.

\begin{figure}[htbp]
\hspace*{-2mm}
\centering
\includegraphics[trim = 0mm 0mm 0mm 0mm, clip=true, angle=0, width=0.8 \columnwidth]{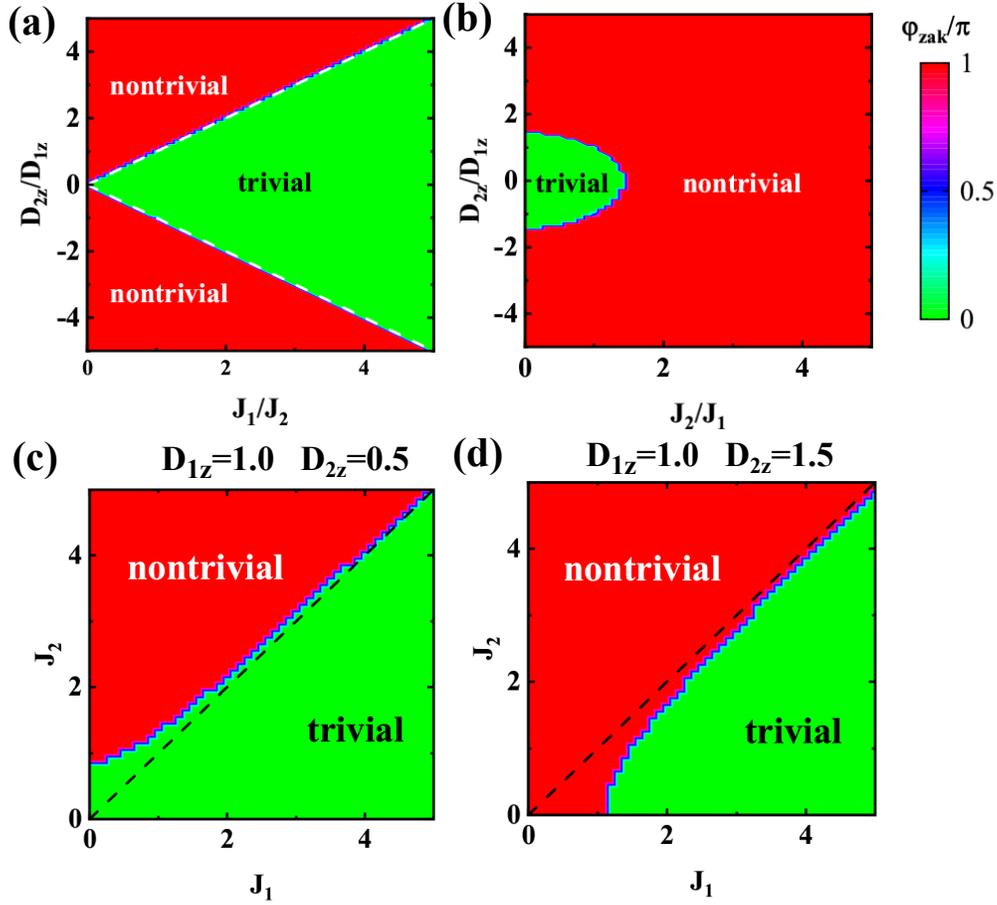}
\caption{(Color online) The $J$-$D$ phase diagram of
$J_{1}/J_{2}$ (a) and $J_{2}/J_{1}$ (b) for the 1D FM SSH model with
the anisotropic DMI. The white dotted lines represent the phase transition
boundaries between the trivial and non-trivial phases given by the relations $\left|D_{2z}/D_{1z}\right|=J_{1}/J_{2}$ ($D_{1z}=J_{2}=1$ is fixed) and $\left(J_{2}/J_{1}\right)^{2}+\left(D_{2z}/D_{1z}\right)^{2}=2$ ($D_{1z}=J_{1}=1$ is fixed), respectively. (c) and (d) are $J_{1}-J_{2}$ phase diagrams at $D_{1z}=1.0>D_{2z}=0.5$ and $D_{1z}=1.0<D_{2z}=1.5$, respectively. The black dashed lines denote the original phase transition boundaries of the pure Heisenberg interactions.}
\label{Fig5}
\end{figure}

\begin{figure}[htbp]
\hspace*{-2mm}
\centering
\includegraphics[trim = 0mm 0mm 0mm 0mm, clip=true, angle=0, width=0.8 \columnwidth]{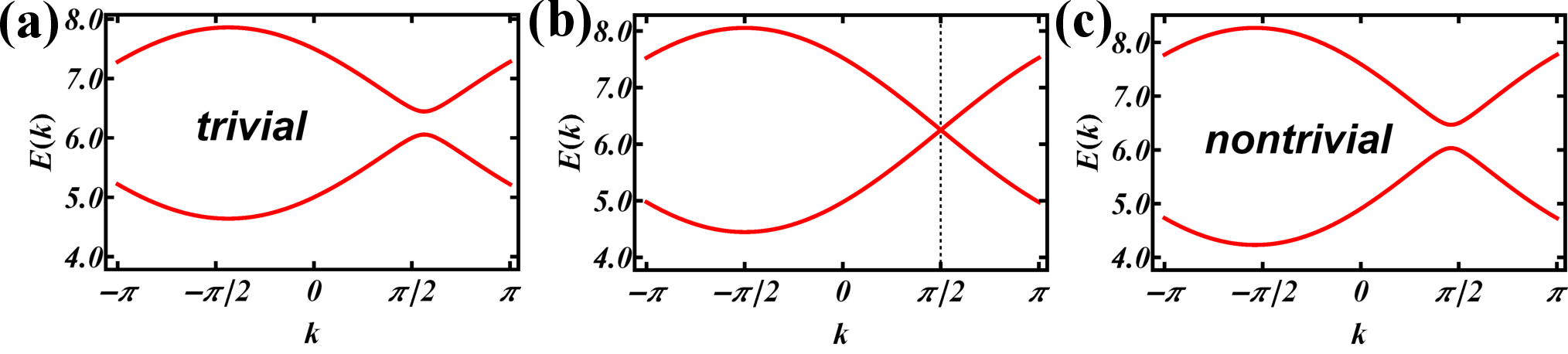}
\caption{(Color online) Band structures dependent on the ratio $D_{2z}/D_{1z}$ at
$J_{1}/J_{2}$=1.5: (a) $D_{2z}/D_{1z}$=1.0 (b) $D_{2z}/D_{1z}$=1.5 (c) $D_{2z}/D_{1z}$=2.0. The vertical black dotted line in (b) indicates the position where the energy gap is closed.}
\label{Fig6}
\end{figure}

\begin{figure}[htbp]
\hspace*{-2mm}
\centering
\includegraphics[trim = 0mm 0mm 0mm 0mm, clip=true, angle=0, width=0.6 \columnwidth]{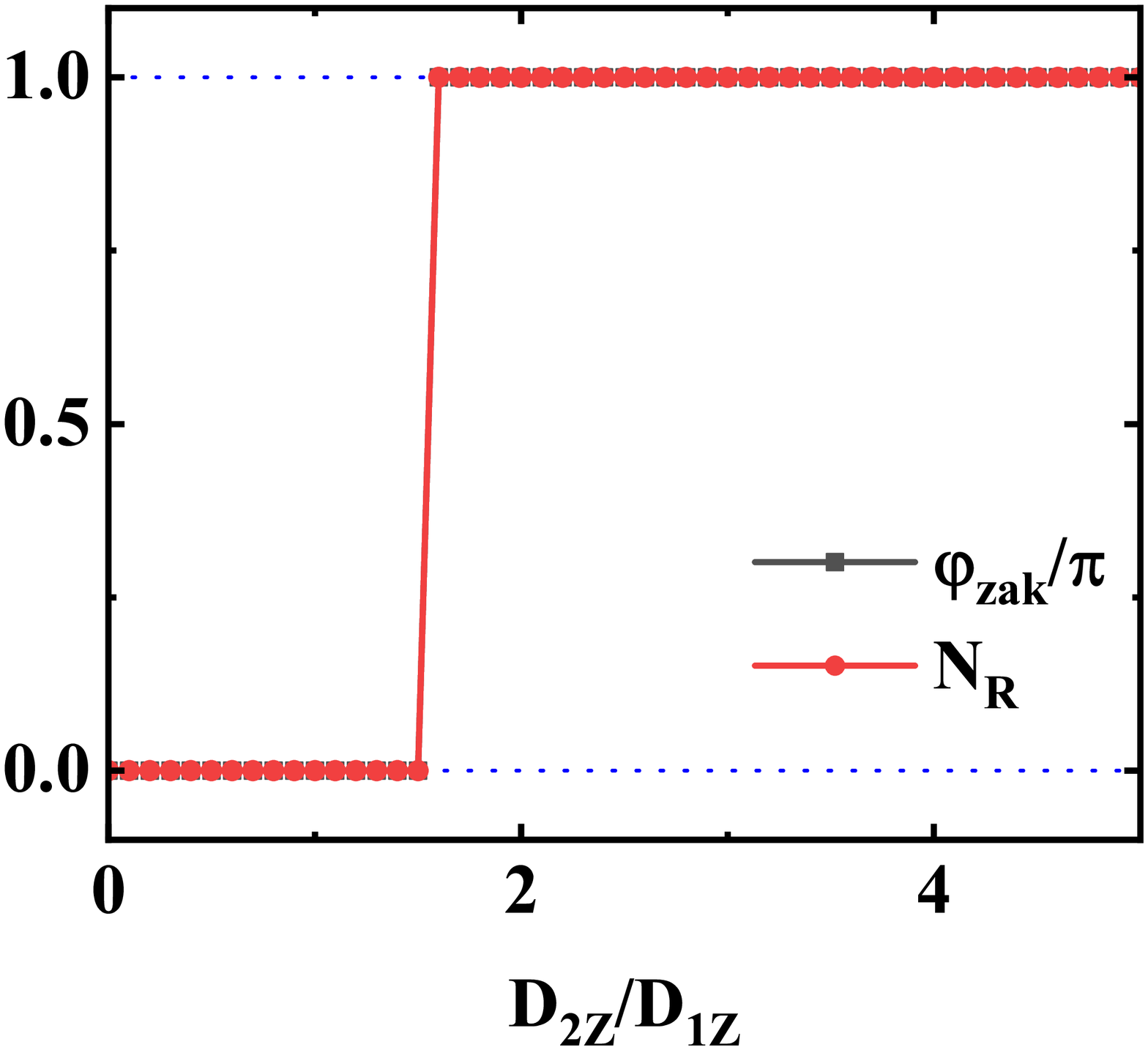}
\caption{(Color online) Momentum $k$-space Zak phase $\varphi_{zak}$ and real-space topological number $N_{R}$ in the 1D FM SSH model with the
Heisenberg and DMI interactions for $J_{1}/J_{2}$=1.5.}
\label{Fig7}
\end{figure}

Compared with the case of pure Heisenberg interaction, the phase
transition condition (more details are shown in Appendix) in the presence of the anisotropic DMI is
$J_{1}^{2}+{D}_{1z}^{2}-J_{2}^{2}-D_{2z}^{2}=0$. Thus, the phase transition occurs at
$\left|D_{2z}/D_{1z}\right|=J_{1}/J_{2}$ for the
fixed $D_{1z}=J_{2}=1$ case, and
$\left(J_{2}/J_{1}\right)^{2}+\left(D_{2z}/D_{1z}\right)^{2}=2$
for the fixed $D_{1z}=J_{1}=1$ case. Obviously, the intercellular
$D_{2}$ (intracellular $D_{1}$) DMI boosts
(weakens) and even induces (eliminates) the topologically non-trivial
phase, though the Heisenberg interaction alone can lead to the
topological magnon phase, which obviously can be seen in Figs.~\ref{Fig5}-\ref{Fig7}.

For the quantization $x$-axis, as displayed in Fig.~\ref{Fig1}(c), the DMI term $H_{DM}^{'}$ can be rewritten as
\begin{eqnarray}
H_{DM}^{'}&=&-iD_{1x}S\sum_{i}\left(a_{i}^{\dag}b_{i}-a_{i}b_{i}^{\dag} \right)\\ \nonumber
&&-iD_{2x}S\sum_{i,i+1}\left(b_{i}^{\dag}a_{i+1}-b_{i}a_{i+1}^{\dag} \right),
  \label{Eq11}
\end{eqnarray}
where $D_{1x}$ and $D_{2x}$ are the components of
$\mathbf{D}_{1}$ and $\mathbf{D}_{2}$ on the quantization $x$-axis of
spin, respectively. From the analysis of the DMI, we can find that only
the components in the direction of quantization axis will affect the
topological properties of the system. Therefore, the conclusions drawn above are still valid.

\subsection{In the presence of PDI}
\begin{figure}[htbp]
\hspace*{-2mm}
\centering
\includegraphics[trim = 0mm 0mm 0mm 0mm, clip=true, angle=0, width=0.8 \columnwidth]{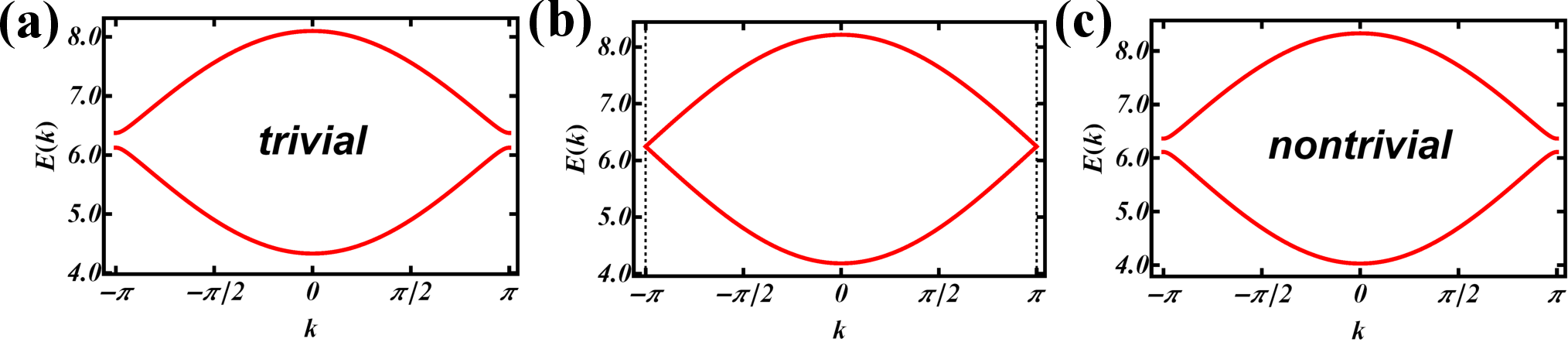}
\caption{(Color online) Band structures dependent on the ratio
$F_{2}/F_{1}$ for $J_{1}/J_{2}$=1.5: $F_{2}/F_{1}$=1.5 (a), 2.0 (b), and  2.5 (c). The vertical black dotted lines in (b) denote the positions of the gap closed for the PDI case, different from that of the DMI case.}
\label{Fig8}
\end{figure}

\begin{figure}[htbp]
\hspace*{-2mm}
\centering
\includegraphics[trim = 0mm 0mm 0mm 0mm, clip=true, angle=0, width=0.8 \columnwidth]{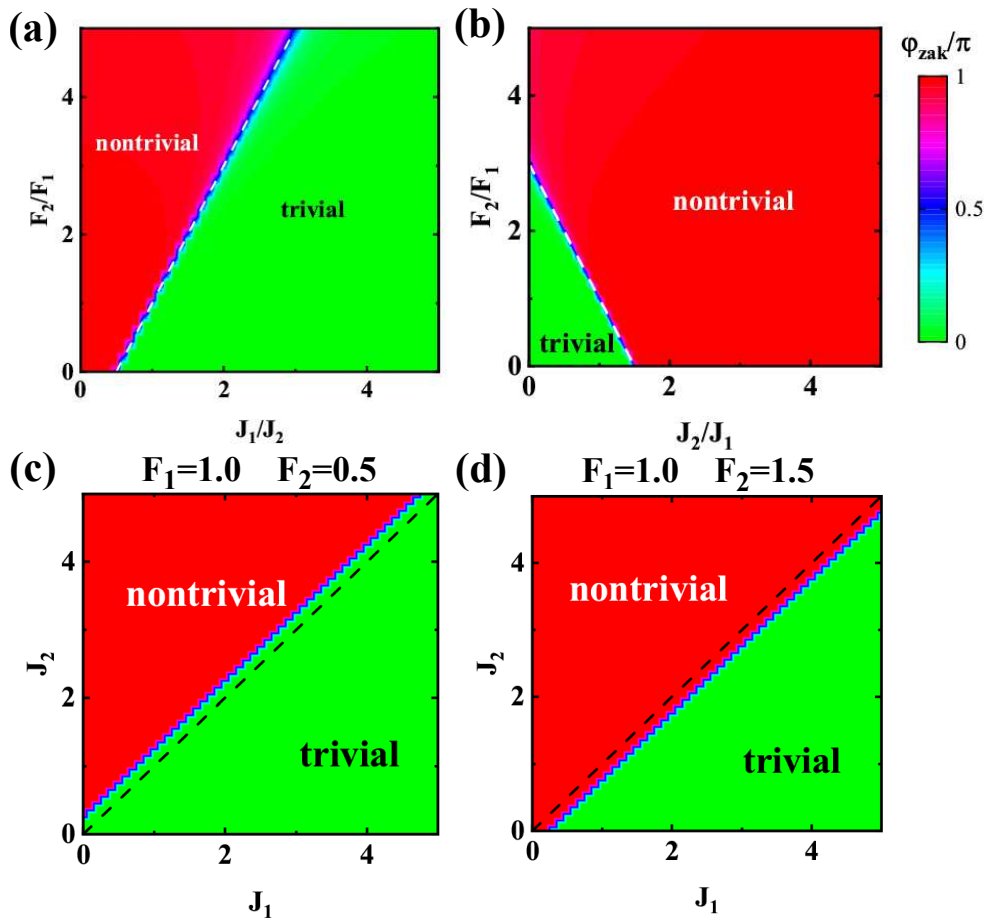}
\caption{(Color online) The $J$-$F$ phase diagram of $J_{1}/J_{2}$ (a) and
$J_{2}/J_{1}$ (b) for the 1D FM SSH model with the anisotropic PDI. The white dotted lines represent the phase transition boundaries between the trivial and non-trivial phases given by the relations $2J_{1}/J_{2}-1=F_{2}/F_{1}$ ($F_{1}=J_{2}=1$ is fixed) and $-2J_{2}/J_{1}+3=F_{2}/F_{1}$ ($F_{1}=J_{1}=1$ is fixed), respectively. (c) and (d) are $J_{1}-J_{2}$ phase diagrams at $F_{1}=1.0>F_{2}=0.5$ and $F_{1}=1.0<F_{2}=1.5$, respectively. The black dashed lines denote the original phase transition boundaries of the pure Heisenberg interactions.}
\label{Fig9}
\end{figure}

\begin{figure}[htbp]
\hspace*{-2mm}
\centering
\includegraphics[trim = 0mm 0mm 0mm 0mm, clip=true, angle=0, width=0.6 \columnwidth]{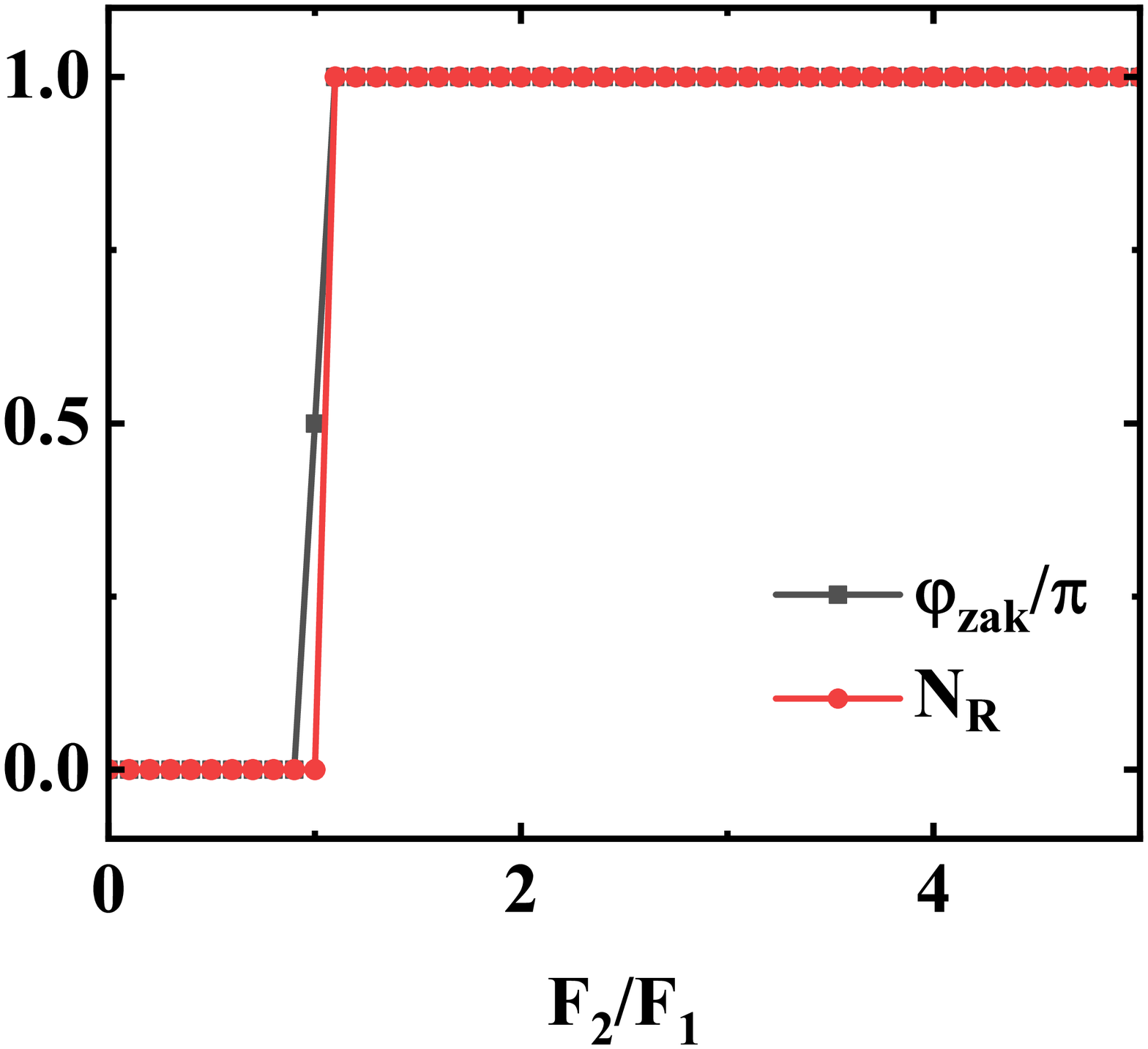}
\caption{(Color online) Momentum $k$-space Zak phase $\varphi_{zak}$ and real-space topological number $N_{R}$ in the 1D FM SSH model with the
Heisenberg and PDI interactions for $J_{2}/J_{1}=1.0$.}
\label{Fig10}
\end{figure}

Apart from the DMI, the PDI is another essential anisotropic exchange interaction, which is generally from the combination of the strong spin-orbit coupling and the orbital degree of freedom \cite{PRL102-017205}. Consequently, the PDI has the different influence on magnons due to its origin distinctly different from that of the DMI. For the quantization $z$-axis, using the linear HP transformation, the PDI term $H_{PD}$ (more details are shown in Appendix) are described as
\begin{eqnarray}
H_{PD}&=&-\frac{F_{1}}{2}S\sum_{i}\left(a_{i}b_{i}+a_{i}b_{i}^{\dag}+ a_{i}^{\dag}b_{i}+a_{i}^{\dag}b_{i}^{\dag} \right) \\ \nonumber
&&-\frac{F_{2}}{2}S\sum_{i,i+1}\left(b_{i}a_{i+1}+b_{i}a_{i+1}^{\dag}+ b_{i}^{\dag}a_{i+1}+b_{i}^{\dag}a_{i+1}^{\dag} \right).
  \label{Eq12}
\end{eqnarray}
In this 1D chain case, the direction of the PDI is along the $x$-axis and
perpendicular to the quantization $z$-axis. For simplicity,
$F_{1}=J_{2}=1$ for the $J_{1}/J_{2}$ case and $F_{1}=J_{1}=1$ for the $J_{2}/J_{1}$ case is fixed in all the
calculations. The phase transition condition (more details are shown in Appendix) in the case of the Heisenberg interaction and PDI is $F_{1}+2J_{1}-F_{2}-2J_{2}=0$. Thus the boundary between the trivial and non-trivial phases is determined by the relation $2J_{1}/J_{2}-1=F_{2}/F_{1}$ for $F_{1}/J_{2}=1$, and $-2J_{2}/J_{1}+3=F_{2}/F_{1}$ for $F_{1}/J_{1}=1$. Similar to the Heisenberg interaction and DMI cases, the intercellular $F_{2}$ (intracellular $F_{1}$) PDI favors (unfavors) the topologically non-trivial phase, as demonstrated in Figs.~\ref{Fig8}-\ref{Fig10}.
Compared with the cases of the Heisenberg and DMI interactions, the position of gap close for the pure PDI case is at $k=\pi$, same as the Heisenberg one. This is a result of the different symmetries of these three magnetic interactions.

For the quantization $x$-axis, as displayed in Fig.~\ref{Fig1}(c), the PDI term $H_{PD}^{'}$ can be rewritten as
\begin{eqnarray}
H_{PD}^{'}&=&-\left(F_{1}+F_{2} \right)S\sum_{i}\left( a_{i}^{\dag}a_{i}+b_{i}^{\dag}b_{i} \right),
  \label{Eq13}
\end{eqnarray}
and then the PDI becomes a term similar to the easy-axial anisotropy $K$ term, equivalent to the on-site energy of magnon. Therefore, the PDI has no influence on the topological properties in this case.

\subsection{In the presence of both DMI and PDI}
Finally, we turn to the case of coexistence of both the anisotropic DMI and PDI. In fact, the Heisenberg interaction, DMI and PDI may possibly coexist in the same realistic magnetic materials due to the the existence of multiple degrees of freedom, especially in 3$d$, 4$d$ and even 5$d$ correlated materials, such as magnetic materials in iridates \cite{PRL105-027204,PRB84-054409}, {\it etc}. However, the phase transition condition is no longer linear. For simplicity, both $D_{1z}=J_{1}$, $F_{1}=J_{1}$ and $J_{1}=1$ are fixed in all the calculations. Then the phase transition boundary can be determined numerically.

\begin{figure}[htbp]
\hspace*{-2mm}
\centering
\includegraphics[trim = 0mm 0mm 0mm 0mm, clip=true, angle=0, width=0.8 \columnwidth]{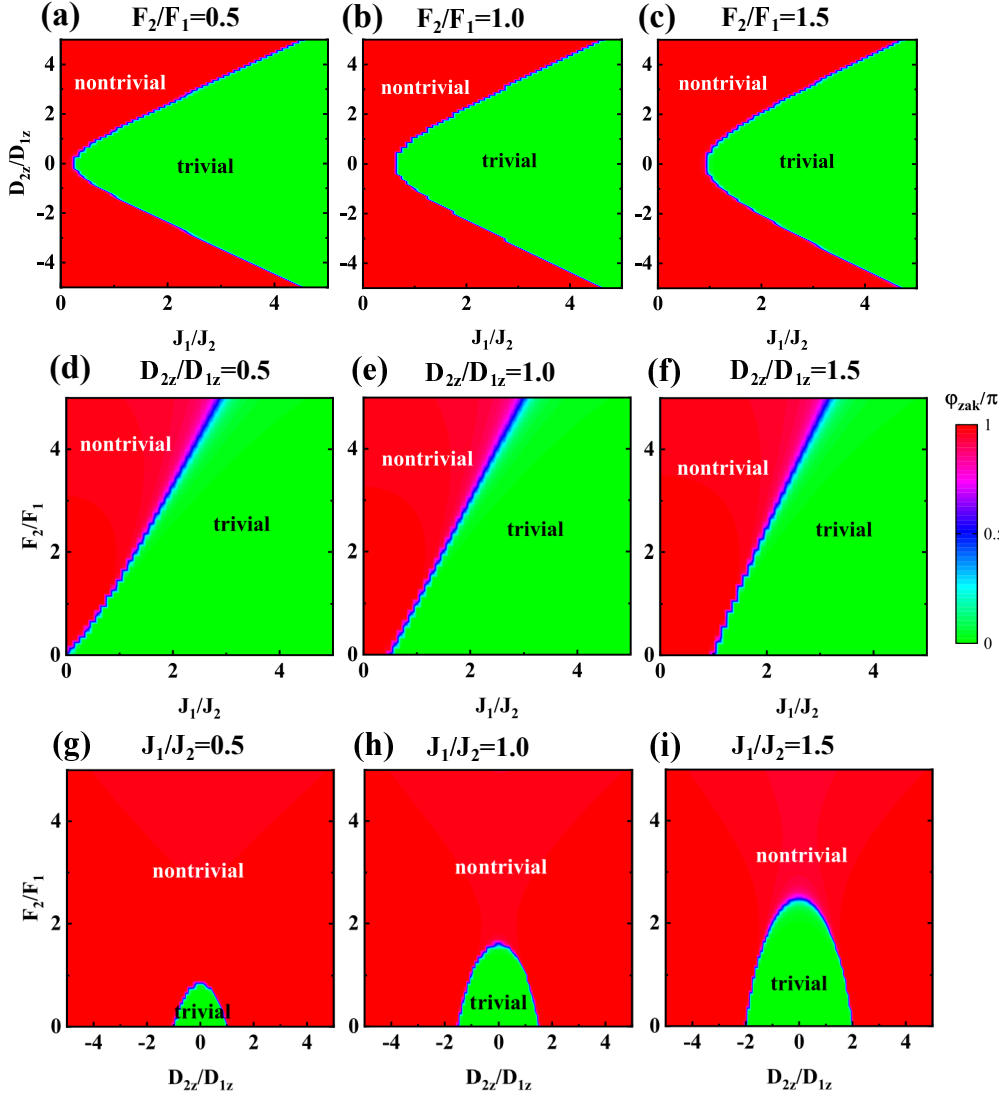}
\caption{(Color online) The $J$-$D$, $J$-$F$ and $D$-$F$ phase diagrams for the 1D FM SSH model with both the anisotropic DMI and PDI.}
\label{Fig11}
\end{figure}

As shown in Figs.~\ref{Fig11}, comparing the $J$-$D$, $J$-$F$ and $D$-$F$ phase diagrams, it is found that the coexistence of the isotropic Heisenberg interaction, anisotropic DMI and PDI show different topological phase diagrams from these in Fig.~\ref{Fig5}(a) and Fig.~\ref{Fig9}(a) as a result of the interplay among them.
Evidently, these three interactions have synergistic effects on the generation of topologically non-trivial phase. In essence, it is the intercellular interactions ($J_{2}$, $D_{2}$ and $F_{2}$) rather than the intracellular interactions ($J_{1}$, $D_{1}$ and $F_{1}$) that contribute to the emergence of the topological phases. As a consequence, these different magnetic exchange interactions introduce a rich topological phase diagram in the 1D magnon system, suggesting a realization of multiple manipulation of topological magnonics. As matter of fact, the relative strength between the isotropic and anisotropic exchange interactions, can be changed by the strain and pressure regulations in the realistic materials, subsequently the topological phase transition of magnons can be achieved.

\section{Remarks and Conclusions}
We notice that different from our local spins, topological magnons in 1D itinerant flatband ferromagnet are also proposed \cite{PRB97-245111}. Indeed, the 1D topological magnons can be explored in realistic one-dimensional (1D) magnets \cite{PRB104-085429}, quasi-1D magnetic materials \cite{NJP15-093043}, or 1D organic magnetic systems \cite{NM17-308,NRM5-87}. As can be expected, more 1D magnetic models rather than the SSH model can be extended to topological magnons. Another way to implement the 1D ferromagnetic SSH model is to strip an edge from the magnonic 2D SSH model in honeycomb ferromagnets \cite{PRB103-014407}.
On the other hand, our results also imply that the topological magnons in the 1D systems are distinctly different from these in the 2D ones, introducing a new platform for the topological magnons.

In summary, we construct a one-dimensional (1D) ferromagnetic (FM) SSH model with the anisotropic interactions, and find a topological magnon non-trivial phase described by the Zak phase, induced not only by the pure strong intercellular isotropic Heisenberg interaction $J_{1}/J_{2}<1$, but also by the strong intercellular anisotropic DMI and PDI. The intercellular anisotropic DMI and PDI in combination with the intercellular isotropic Heisenberg-type interaction manifest a synergistic effect on the topological phase transition. In addition, the quantization axis of spin also substantially affect the topological magnon phase diagram owing to the anisotropic interactions. Of particular interest, the existence of topologically protected in-gap edge states (magnon bound states) in 1D magnets with FM SSH model provides a possible route to Majorana-like particle realization. Due to the rich manipulated magnetic interactions and the 1D structural characteristics, abundant topological magnon states and magnonic crystals can be designed from bottom to top, suggesting potential applications in the field of topological magnonics.

\acknowledgements
This work was supported by the National Sciences Foundation of China under Grant Nos. 11974354, 11774350 and 11574315. The calculations were performed in Center for Computational Science of CASHIPS, the ScGrid of Supercomputing Center and Computer Network Information Center of Chinese Academy of Science.

\section{Appendix}
For the quantization $z$-axis, the explicit matrix form of $H_{0}$,
$H_{DM}$ and $H_{PD}$ are
\begin{eqnarray}
H_{0}=\begin{pmatrix}
\varepsilon_{0} & 0 & h(k) & 0 \\
0 & \varepsilon_{0} & 0 & h^{*}(k) \\
h^{*}(k) & 0 & \varepsilon_{0} & 0 \\
0 & h(k) & 0 & \varepsilon_{0}
\end{pmatrix}
\end{eqnarray}
\begin{eqnarray}
H_{DM}=\begin{pmatrix}
0 & 0 & f(k) & 0 \\
0 & 0 & 0 & f^{*}(k) \\
f^{*}(k) & 0 & 0 & 0 \\
0 & f(k) & 0 & 0
\end{pmatrix}
\end{eqnarray}
\begin{eqnarray}
H_{PD}=\begin{pmatrix}
0 & 0 & g(k) & g(k) \\
0 & 0 & g^{*}(k) & g^{*}(k) \\
g^{*}(k) & g(k) & 0 & 0 \\
g^{*}(k) & g(k) & 0 & 0
\end{pmatrix}
\end{eqnarray}
where $\varepsilon_{0}=\left(J_{1}+J_{2}+K \right)S$, $h(k)=-J_{1}S- J_{2}Se^{-ika}$, $f(k)=-iD_{1}S+iD_{2}Se^{-ika}$, $g(k)=\left(-F_{1}S-F_{2}Se^{-ika}\right)/2$.

For $H_{1}=H_{0}+H_{DM}$, we assume $J_{1}/J_{2}=\alpha>0$,
$D_{1z}/J_{2}=\beta_{1}>0$, $D_{2z}/D_{1z}=\gamma_{1}$,
$D_{2z}/J_{2}=\gamma_{1}\beta_{1}$. When the topological phase transition occurs at the closed energy gap, and the parameters are satisfied by
\begin{eqnarray}
\begin{cases}
-\alpha-\cos(ka)+\gamma_{1}\beta_{1}\sin(ka)=0 \\
\sin(ka)-\beta_{1}+\gamma_{1}\beta_{1}\cos(ka)= 0 \\
\end{cases}
\end{eqnarray}

Cancelling $sin(ka)$ and $cos(ka)$, we find $\beta_{1}^{2}+\alpha^{2}=1+\gamma_{1}^{2}\beta_{1}^{2}$. Here
we take $\beta_{1}=1$ and $a=1$. As a result, the condition of
the phase transition ({\it i.e.} the gap closed) is $\alpha^{2}=\gamma_{1}^{2}$. When $\gamma_{1}=\alpha$, $\sin(ka)=1$, that is, $k=\pi/2a$; when $\gamma_{1}=-\alpha$, $\sin(ka)=(1-\alpha^{2})/(1+\alpha^{2})$, $\cos(ka)=-2\alpha/(1+\alpha^{2})$.

For $H_{2}=H_{0}+H_{PD}$, we assume $F_{1}/J_{2}=\beta_{2}>0$,
$F_{2}/F_{1}=\gamma_{2}>0$, $F_{2}/J_{2}=\gamma_{2}\beta_{2}$. The parameters are satisfied by
\begin{eqnarray}
(2\alpha+\beta_{2})^{2}+(2+\gamma_{2}\beta_{2} )^{2}+2(2\alpha+\beta_{2})(2+\gamma_{2}\beta_{2} )\cos(ka)=0
\end{eqnarray}
Obviously, $-1\leq \cos(ka)<0$. Assuming $n=\cos(ka)-(-1)$, it can be rewritten as
\begin{eqnarray}
((2\alpha+\beta_{2})-(2+\gamma_{2}\beta_{2} ))^{2}+2n(2\alpha+\beta_{2})(2+\gamma_{2}\beta_{2} )=0
\end{eqnarray}
We can see that it is valid only for $n=0$, {\it i.e.} $\cos(ka)=-1$.
We take $\beta_{2}=1$, and then $\gamma_{2}=2\alpha-1$. Then the condition of topological phase transition can be determined analytically.

\end{document}